\documentclass[prd,twocolumn,twoside,preprintnumbers,superscriptaddress,nofootinbib]{revtex4}

\usepackage{amsmath,slashed}
\usepackage{graphicx,graphics,color}
\usepackage{dcolumn}
\usepackage[hyperfootnotes=false]{hyperref}
\usepackage{xspace}
\usepackage{enumerate}
\usepackage{varwidth}
\usepackage[normalem]{ulem}

\usepackage{physics}
\usepackage{amssymb}
\usepackage{mathtools}
\usepackage{dsfont}
\usepackage{multirow}

\hypersetup{
    colorlinks=true,
    linkcolor=blue,
    citecolor=blue
}

\newcommand{\GeV}{\,\text{GeV}}

\newcommand{\TeV}{\,\text{TeV}}

\renewcommand{\Im}{\text{Im}\,}
\renewcommand{\Re}{\text{Re}\,}

\allowdisplaybreaks[1]

\begin{document}

\title{Light new physics and the $\boldsymbol{\tau}$ lepton dipole moments: prospects at Belle II}

\author{Martin Hoferichter}
\affiliation{Albert Einstein Center for Fundamental Physics, Institute for Theoretical Physics, University of Bern, Sidlerstrasse 5, 3012 Bern, Switzerland}

\author{Gabriele Levati}
\affiliation{Albert Einstein Center for Fundamental Physics, Institute for Theoretical Physics, University of Bern, Sidlerstrasse 5, 3012 Bern, Switzerland}

\begin{abstract} 
While electron and muon dipole moments are well-established precision probes of physics beyond the Standard Model, it is notoriously challenging to test realistic New-Physics (NP) scenarios for the $\tau$ lepton. Constructing suitable asymmetries in $e^+e^-\to\tau^+\tau^-$ has emerged as a promising such avenue, providing access to the electric and magnetic dipole moment once a polarized electron beam is available, e.g., with the proposed polarization upgrade of the SuperKEKB $e^+e^-$ collider. However, this interpretation relies on an effective-field-theory (EFT) argument that only applies if the NP scale is large compared to the center-of-mass energy. In this Letter we address the consequences of the asymmetry measurements in the case of light NP, using 
light spin-0 and spin-1 bosons as test cases, to show how results can again be interpreted as constraints on dipole moments, albeit in a model-dependent manner, and how the decoupling to the EFT limit proceeds in these cases. In particular, we observe that the imaginary parts generated by light new particles can yield nonvanishing asymmetries even without electron polarization, which can again be interpreted as constraints on the $\tau$ anomalous magnetic moment. This proposed measurement thus presents a novel opportunity
for NP searches that can be realized already with present data at Belle II.
\end{abstract}

\maketitle

\emph{Introduction}---Lepton dipole moments constitute some of the most interesting observables to test the Standard Model (SM) of particle physics. Indeed, the electric ($d_\ell$) and magnetic ($a_\ell$) dipole moments of electrons and muons are low-energy precision observables that have been experimentally measured or constrained to extremely high precision:
\begin{align}
\label{dipole_moments_exp}
a_\mu^{\text{exp}} &= 116\, 592\, 071.5(14.5) \times
10^{-11} &\text{\cite{Muong-2:2025xyk}}\,, \nonumber \\
a_e^{\text{exp}} &= 115 \, 965\,  218 \, 059\,(13) \times
10^{-14}  &\text{\cite{Fan:2022eto}}\,,\nonumber\\
d_e^\text{exp} &< 4.1 \times 10^{-30} \, e\, \text{cm}\,&\text{\cite{Roussy:2022cmp}}\,,\nonumber\\
d_\mu^\text{exp} &< 2 \times 10^{-19} \, e\, \text{cm}\,&\text{\cite{Muong-2:2008ebm}}\,.
\end{align}
Among these New Physics (NP) probes,
$d_\mu$ is clearly the least explored one, with several efforts under way to improve the precision~\cite{Abe:2019thb,Aiba:2021bxe,Adelmann:2025nev,Crivellin:2018qmi},
but even for $d_e$ the SM expectation is still away by about five orders of magnitude~\cite{Ema:2022yra} (and many more for $d_{\mu,\tau}$~\cite{Yamaguchi:2020eub}).
For $a_\ell$, commensurate efforts in experiment and theory are required to fully leverage the NP sensitivity. For $a_e$, it is a persistent tension between measurements of the fine-structure constant in Cs~\cite{Parker:2018vye} and Rb~\cite{Morel:2020dww} atom-interferometry experiments that currently limits the reach, while theory uncertainties~\cite{Aoyama:2019ryr,Volkov:2019phy,Volkov:2024yzc,Aoyama:2024aly,DiLuzio:2024sps,Hoferichter:2025fea} at present are a factor of four below the uncertainty of the direct measurement~\cite{Fan:2022eto}. In contrast, the global average of $a_\mu^\text{exp}$~\cite{Muong-2:2025xyk,Muong-2:2023cdq,Muong-2:2024hpx,Muong-2:2021ojo,Muong-2:2021vma,Muong-2:2021ovs,Muong-2:2021xzz,Muong-2:2006rrc} is currently a factor of four more precise than the theory prediction~\cite{Aliberti:2025beg,Aoyama:2012wk,Volkov:2019phy,Volkov:2024yzc,Aoyama:2024aly,Parker:2018vye,Morel:2020dww,Fan:2022eto,Czarnecki:2002nt,Gnendiger:2013pva,Ludtke:2024ase,Hoferichter:2025yih,RBC:2018dos,Giusti:2019xct,Borsanyi:2020mff,Lehner:2020crt,Wang:2022lkq,Aubin:2022hgm,Ce:2022kxy,ExtendedTwistedMass:2022jpw,RBC:2023pvn,Kuberski:2024bcj,Boccaletti:2024guq,Spiegel:2024dec,RBC:2024fic,Djukanovic:2024cmq,ExtendedTwistedMass:2024nyi,MILC:2024ryz,FermilabLatticeHPQCD:2024ppc,Keshavarzi:2019abf,DiLuzio:2024sps,Kurz:2014wya,Colangelo:2015ama,Masjuan:2017tvw,Colangelo:2017qdm,Colangelo:2017fiz,Hoferichter:2018dmo,Hoferichter:2018kwz,Eichmann:2019tjk,Bijnens:2019ghy,Leutgeb:2019gbz,Cappiello:2019hwh,Masjuan:2020jsf,Bijnens:2020xnl,Bijnens:2021jqo,Danilkin:2021icn,Stamen:2022uqh,Leutgeb:2022lqw,Hoferichter:2023tgp,Hoferichter:2024fsj,Estrada:2024cfy,Deineka:2024mzt,Eichmann:2024glq,Bijnens:2024jgh,Hoferichter:2024vbu,Hoferichter:2024bae,Holz:2024lom,Holz:2024diw,Cappiello:2025fyf,Colangelo:2014qya,Blum:2019ugy,Chao:2021tvp,Chao:2022xzg,Blum:2023vlm,Fodor:2024jyn}, and substantial efforts are being invested~\cite{Colangelo:2022jxc,Aliberti:2025beg,Hertzog:2025ssc} to realize the NP reach set by experiment.

These tests in $a_\ell$ and $d_\ell$ are highly complementary, probing the flavor and $CP$ properties of potential NP scenarios, e.g., the relative size of NP effects in $a_\ell$ can scale linearly or quadratically with the lepton mass depending on the origin of the chirality flip. From this perspective, it is unfortunate that the short lifetime of the $\tau$ lepton renders its dipole moments much more challenging to access in experiment at a similar level of precision. Apart from discerning chirally enhanced cases~\cite{Giudice:2012ms,Crivellin:2021rbq,Athron:2025ets}, $\tau$ dipole moments would also help test NP scenarios predicting larger couplings to the third generation of fermions~\cite{Barbieri:1995uv, Barbieri:1996ae, Barbieri:2012uh, Matsedonskyi:2014iha,Panico:2016ull,Glioti:2024hye,Froggatt:1978nt,Berezhiani:1983hm,Bordone:2017bld,Fuentes-Martin:2022xnb, Davighi:2023iks,Nakai:2025dmp}.
Intense research programs are currently active in devising new experiments and techniques to precisely measure $a_\tau$ and $d_\tau$, see, e.g., Refs.~\cite{Kim:1982ry,Samuel:1990su,Eidelman:2016aih,Koksal:2017nmy,Fomin:2018ybj,Fu:2019utm,Gutierrez-Rodriguez:2019umw,Beresford:2019gww,Dyndal:2020yen,Haisch:2023upo,Shao:2023bga,Beresford:2024dsc,Dittmaier:2025ikh,Buttazzo:2026amk}, including recent measurements in peripheral Pb--Pb collisions at LHC~\cite{ATLAS:2022ryk,CMS:2022arf,CMS:2024qjo}. Unfortunately, it appears challenging to scale these techniques to a sensitivity much beyond the Schwinger term (except for future lepton colliders~\cite{Buttazzo:2026amk}), while testing realistic NP scenarios requires a precision of at least $10^{-5}$ in $a_\tau$~\cite{Crivellin:2021spu}.

A promising way around these limitations proceeds via suitably chosen asymmetries in $e^+e^-\to\tau^+\tau^-$, as first proposed in Refs.~\cite{Bernabeu:2004ww,Bernabeu:2006wf,Bernabeu:2007rr, Bernabeu:2008ii} at the lower $\Upsilon$ resonances. More recently, the requirements for a practical implementation were studied~\cite{Crivellin:2021spu,USBelleIIGroup:2022qro,Aihara:2024zds}, including the calculation of radiative corrections and development of Monte-Carlo tools~\cite{Gogniat:2025eom,Banerjee:2020rww,Ulrich:2025fij}.
The quantities that are experimentally accessible from these asymmetries are the form factors $F_2(s)$ and $F_3(s)$ of the electromagnetic $\tau \tau \gamma$ vertex at the squared center-of-mass (CM) energy $s$. As the $\tau$ dipole moments represent the zero-momentum counterparts of such form factors, $a_\tau \propto F_2(0)$ and $d_\tau \propto F_3(0)$, a proper subtraction of momentum-dependent corrections has to be performed before drawing conclusive bounds on potential NP effects. For heavy NP candidates, the connection becomes straightforward via an effective-field-theory (EFT) argument, i.e., as long as
the mass scale $m_\text{NP}$ of the NP fields is larger than the energy available at the experiment under consideration ($m_\text{NP}^2\gg s$), they effectively decouple and model-independent conclusions on $a_\tau^{\text{NP}}$ can be drawn.

On the other hand, if NP states are light, i.e.,  they  still propagate at collider energies, $m_\text{NP}^2 \lesssim s$, model-dependent NP contributions to the form factors $F_i(s)$ arise and must be taken into account before drawing any conclusion about $a_\tau$ and $d_\tau$. Given the continuously growing attention that light NP candidates have been receiving in recent years, it is worthwhile to thoroughly investigate such a possibility for the most relevant classes of light NP mediators. These include spin-$0$ particles such as axions, axionlike particles, and generic light scalars, as well as spin-$1$ states such as light vector bosons. In this Letter, we perform such an analysis, with special focus on the decoupling once the mass is increased in these NP scenarios, and express the resulting bounds in each model again in terms of $\tau$ dipole moments. In particular, we will see that imaginary parts that are generated for light NP can be accessed via asymmetries that do not require a polarized electron beam, presenting a novel opportunity for NP searches at $e^+e^-$ colliders such as Belle II.

\emph{Asymmetries in $e^+e^-\to\tau^+\tau^-$}---The general parameterization of the $\gamma \ell \ell$ electromagnetic vertex reads
\begin{align}
\langle \ell(p')|j_{\text{em}}^\mu|\ell(p)\rangle &=  \,e \,\bar{u}(p')\, \Big[\gamma^\mu F_1 + (i F_2 + F_3 \gamma_5)\, \frac{\sigma^{\mu\nu}q_\nu}{2m_\ell}\nonumber \\
& \qquad + \left(q^2 \gamma^\mu - q^\mu \slashed{q}\right)\gamma_5 F_A \Big]\, u(p)\,,
\end{align}
where $q = p'-p$ is the momentum carried by the photon, and the form factors depend on $q^2=s$. $F_1$ describes the vectorial component of the electromagnetic vertex, while $F_A$ encodes the anapole moment.
In the $s\to 0$ limit the form factors $F_2$ and $F_3$ are in direct relation to $a_\ell$ and $d_\ell$:
\begin{equation}
a_\ell = \Re F_2(0) \,, \qquad d_\ell = \frac{e}{2m_\ell} \Re F_3(0)\,.
\end{equation}
In the presence of NP contributions, form factors can be generally parameterized as consisting of a SM contribution and of a NP one,
\begin{equation}
F_{2,3}(s) = F^\text{SM}_{2,3}(s) + F^\text{NP}_{2,3}(s)\,.
\end{equation}
If all the NP states involved in the loop contributions are heavy, i.e., if for each of them $ m_\text{NP}^2\gg s$, they can be integrated out of the theory. Their impact on form factors is necessarily local and real, and it can be parameterized in terms of appropriate SMEFT dipole operators. For these SMEFT-like contributions then $F_2^{\text{NP}}(s/m_\text{NP}^2\ll 1) \simeq a_\tau^\text{NP}$ and $F_3^{\text{NP}}(s/m_\text{NP}^2\ll 1) \simeq 2m_\tau/e\,d_\tau^\text{NP}$.
Hence, for heavy NP in the SMEFT sense we must construct asymmetries that isolate exclusively the real part of form factors, $\Re F_{2,3}$.

\begin{figure}[t]
\centering
\includegraphics[width=\linewidth]{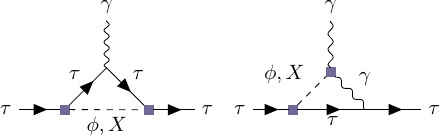}
\caption{Representative Feynman diagrams contributing to $a_\tau$. Square dots denote insertions of NP couplings, $\phi$ and $X$ refer to new spin-$0$ and spin-$1$ particles, respectively.}
\label{fig:RepFeynmanDiagrams}
\end{figure}

If NP is light, instead, the real part of form factors cannot be directly related to the $\tau$ dipole moments. Moreover, its contributions to form factors can develop an imaginary part, provided that $s > (m_i+m_j)^2$, where $m_{i,j}$ are the masses of any two of the particles involved in the loop, see Fig.~\ref{fig:RepFeynmanDiagrams}. Both the real and imaginary parts of such NP contributions do, however, depend on the same NP couplings that are responsible for generating $a_\tau^\text{NP}$ and $d_\tau^\text{NP}$, so that these quantities can be related once the model-specific momentum-dependent contribution to the loop is under control.

\begin{figure*}[t]
    \centering
    \includegraphics[width=0.48\linewidth]{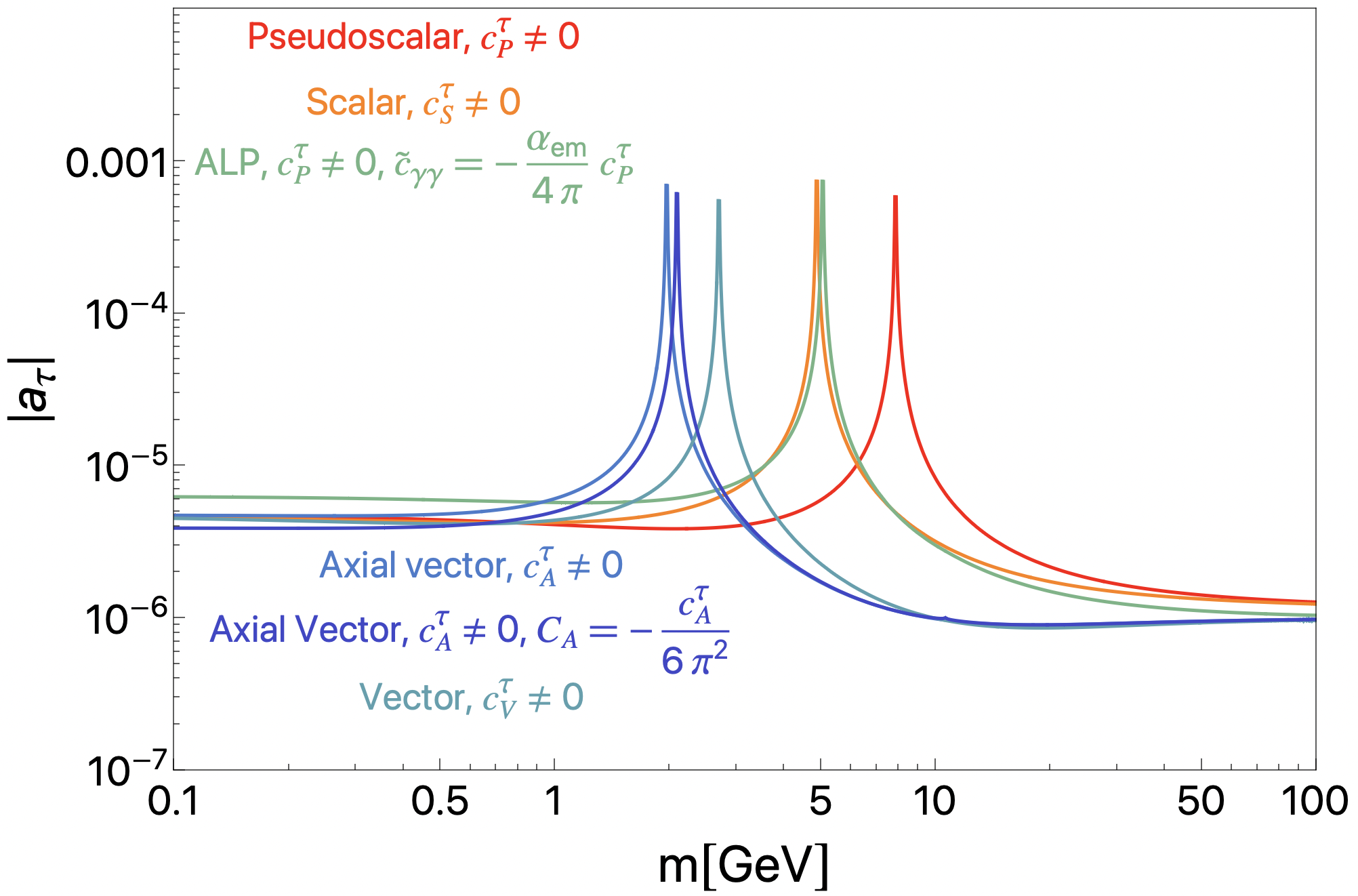}
    \includegraphics[width=0.49\linewidth]{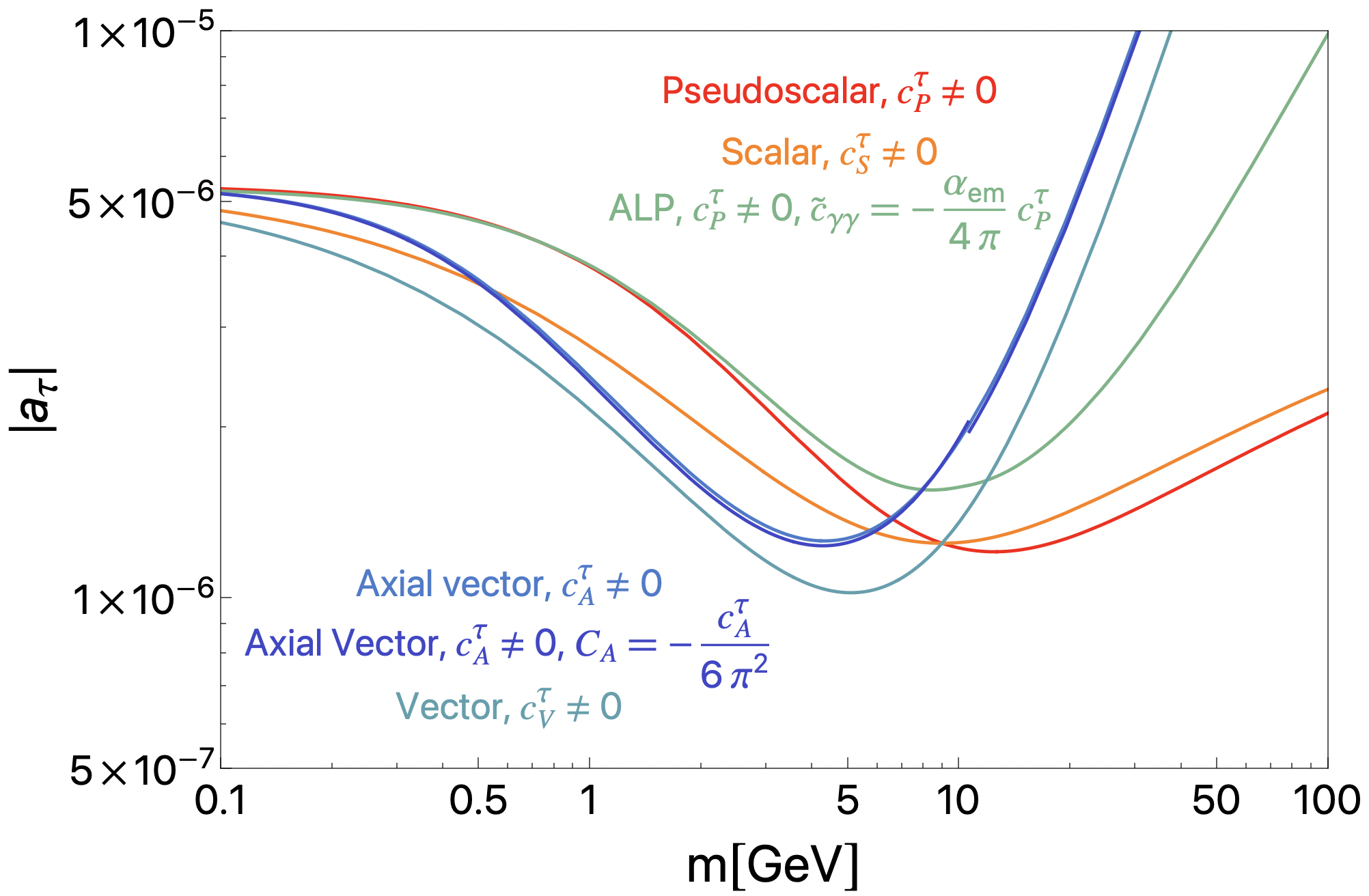}
    \caption{Comparison of the sensitivity of the Belle II experiments to light NP affecting $a_\tau$ for $\Lambda = 1\TeV$. The choice of $c_{\gamma\gamma} = -\alpha_\text{em}/(4\pi) \,c_P^\tau$ corresponds to the minimal coupling present in the case of a derivatively coupled axionlike particle, being unavoidably generated in passing from the derivative to the nonderivative basis. The choice of $C_A = c_A^\tau/(6\pi^2)$ is dictated by the requirement of gauge anomaly cancellation. \textit{Left (Right)}: Bounds obtained assuming a sensitivity on $\Re F_2^\text{eff} (\Im F_2^\text{eff})= 10^{-6}$ as a function of the mass of the NP mediator. The figures indicate the NP reach for a given sensitivity in $\Re F_2^\text{eff} (\Im F_2^\text{eff})$, whose extraction needs to account for radiative corrections, see Ref.~\cite{Gogniat:2025eom}.}
    \label{fig:Money_Plots}
\end{figure*}

Access to $\Re F_2$ and $\Re F_3$ can be gained provided that polarized electron beams are available for the process $e^+ e^- \to \tau^+ \tau^-$, reconstructing $\tau^\pm$ via semileptonic decays into a hadron $h^\pm$. Following the notation of Ref.~\cite{Gogniat:2025eom}, the required asymmetries are $A_T^\pm$, $A_L^\pm$, and $A_{N,F_3}^\pm$, from which $\Re F_{2,3}$ can be inferred via
\begin{align}
\label{eq:F2F3eff_anomalies}
\Re  F_2^\text{eff} &\equiv \mp \frac{4s\beta_e\sigma_\text{tot}}{\pi^2\alpha^2\beta_\tau^3\gamma_\tau \alpha_\pm} \left(A_T^\pm - \frac{\pi}{2\gamma_\tau}A_L^\pm\right)\notag\\
&=\Re (F_2F_1^*)+ |F_2|^2\,,\notag\\
\Re  F_3^\text{eff} &\equiv \frac{4s\beta_e\sigma_\text{tot}}{\pi^2\alpha^2\beta_\tau^2\gamma_\tau \alpha_\pm}A_{N, F_3}^\pm\notag\\
&=\Re (F_3F_1^*)+ \Re (F_3F_2^*)\,,
\end{align}
where we have defined the kinematic variables
\begin{equation}
 \beta_\ell = \sqrt{1-\frac{4m_\ell^2}{s}}\,, \qquad \gamma_\ell = \frac{\sqrt{s}}{2m_\ell}\,,
\end{equation}
and $\alpha_\pm$ denotes the polarization analyzer of the hadronic state.  Referring to Ref.~\cite{Gogniat:2025eom} for explicit expressions, $A_T^\pm$, $A_L^\pm$, and $A_{N,F_3}^\pm$ are then defined via asymmetries involving the scattering angle, the decay angles of $\tau^\pm\to h^\pm\nu_\tau$, and the polarization of the incoming electron.
The identifications in Eq.~\eqref{eq:F2F3eff_anomalies} hold true when considering form-factor-type diagrams, i.e., assuming that more complicated radiative corrections have been removed in the analysis, see Ref.~\cite{Gogniat:2025eom}. In this way, a measurement of $\Re  F_2^\text{eff}$ and $\Re  F_3^\text{eff}$ along these lines provides access to the sought interference terms, and for heavy NP it follows\footnote{$\Re F_3^\text{eff}$ and $\Im F_3^\text{eff}$, defined in Eq.~\eqref{ImF3eff},  do not possess definite $CP$-transformation properties and hence do receive a small contribution from $\gamma$--$Z$ interference in the SM, which can either be subtracted or removed by an average of the $h^\pm$ results~\cite{Bernabeu:2004ww,Bernabeu:2006wf}.}
\begin{align}
 a_\tau^\text{NP} &= \Re F_2^\text{eff}\Big|_\text{exp}-\Re F_2^\text{eff}\Big|_\text{SM}\,,\notag\\
 d_\tau^\text{NP} &= \frac{e}{2m_\tau}\Re F_3^\text{eff}\Big|_\text{exp}\,.
\end{align}

While, in the asymmetry-based approach, $\Re F_{2,3}$ both require electron polarization, $\Im  F_2$ can be measured also in the absence of polarized electron beams. Indeed, it is sufficient to construct the following effective quantity,
\begin{equation}
\Im F_2^\text{eff} \equiv \pm \frac{3s\sigma_\text{tot}}{\pi\alpha^2\beta_e\beta_\tau^3\gamma_\tau\alpha_\pm}A_N^\pm
=\Im (F_2F_1^*)\,,
\end{equation}
where the normal asymmetry $A_N^\pm$ only requires access to the polarization of the $\tau$ decay products and the scattering angle, see Ref.~\cite{Gogniat:2025eom} for more details. Similarly, one can construct an asymmetry that projects out the imaginary part of $F_3$, by means of
\begin{align}
\label{ImF3eff}
 \Im F_3^\text{eff} &\equiv \frac{3s\sigma_\text{tot}}{\pi\alpha^2\beta_e\beta_\tau^2\gamma_\tau\alpha_\pm}A_{T,F_3}^\pm\notag\\
&=\Im (F_3F_1^*)+\Im (F_3F_2^*)\,,
\end{align}
where $A_{T,F_3}^\pm$ is defined in analogy to $A_T^\pm$ with the helicity difference $\sigma_\text{pol}$ replaced by the forward--backward asymmetry $\sigma_\text{FB}$~\cite{Gogniat:2025eom,Bernabeu:2004ww}. Accordingly, $\Im F_3$ is also accessible via asymmetries that do not require electron polarization.

\emph{Light New Physics scenarios}---Among the best motivated light NP scenarios that are currently being investigated are light spin-$0$ particles (axions~\cite{Peccei:1977ur, Peccei:1977hh,Weinberg:1977ma, Wilczek:1977pj}, axionlike particles~\cite{Jaeckel:2010ni, Preskill:1982cy, Abbott:1982af, Dine:1982ah, Davidson:1981zd, Wilczek:1982rv, Graham:2015cka}, dilatons~\cite{Salam:1969bwb, Ellis:1970yd, Goldberger:2007zk}) and light vector bosons~\cite{Langacker:1980js, Hewett:1988xc,Antoniadis:1990ew,Cvetic:1997mbh,Fayet:2007ua,Langacker:2008yv,Dror:2017nsg, Dror:2018wfl}.
In order to identify the contributions of such classes of models to the $\tau$ dipole moments we adopt an EFT approach and parameterize their interactions with $\tau$ leptons and photons limiting ourselves to $U(1)_\text{em}$-invariant and flavor-diagonal couplings.\footnote{We disregard weak interactions because they are expected to play a subleading role at the operational energies of Belle II~\cite{Gogniat:2026zvf}. Indeed, weak corrections will experience a suppression of the order of a few percent, $s_B/M_{W,Z}^2 \simeq 0.01$.}

For a scalar $\phi$ we consider the following interactions,
\begin{align}
    \label{eq:ALP_Lag}
            \mathcal{L}^\text{int}_{\phi} &= \phi \frac{m_\tau}{\Lambda}\,\bar{\tau}\left(c_S^\tau + i \,c_P^\tau \,\gamma_5\right) \tau \\
            & \qquad+   c_{\gamma\gamma} \frac{\alpha_\text{em}}{4\pi}\frac{\phi}{\Lambda}F_{\mu\nu}F^{\mu\nu}+  \, \tilde{c}_{\gamma\gamma} \frac{\alpha_\text{em}}{4\pi}\frac{\phi}{\Lambda}F_{\mu\nu}\tilde{F}^{\mu\nu}\,,\notag
\end{align}
whereas for a vector $X_\mu$ we consider the following set of interactions, allowing for a possible anomalous Chern--Simons term:
\begin{align}
    \label{eq:XBos}
            \mathcal{L}^\text{int}_{X} &= i \, g_D\, X_\mu \,\bar{\tau}\, \gamma^\mu \left(c_V + \,c_A \,\gamma_5\right) \tau \notag\\
            &\qquad+ g_D \, e^2\, C_{A} \,\varepsilon^{\mu\nu\alpha\beta}X_\mu A_\nu \partial_\alpha A_\beta \,.
\end{align}
Given this set of interactions, we then compute the contributions of such new states to the dipole moments $a_\tau$ and $d_\tau$, as well as their contributions to form factors $F_{2,3}(q^2)$, see Fig.~\ref{fig:RepFeynmanDiagrams} for a representative set of diagrams. Explicit expressions are provided in Ref.~\cite{Hoferichter:2025zjp}.

\begin{figure*}[t]
    \centering
    \includegraphics[width=0.48\linewidth]{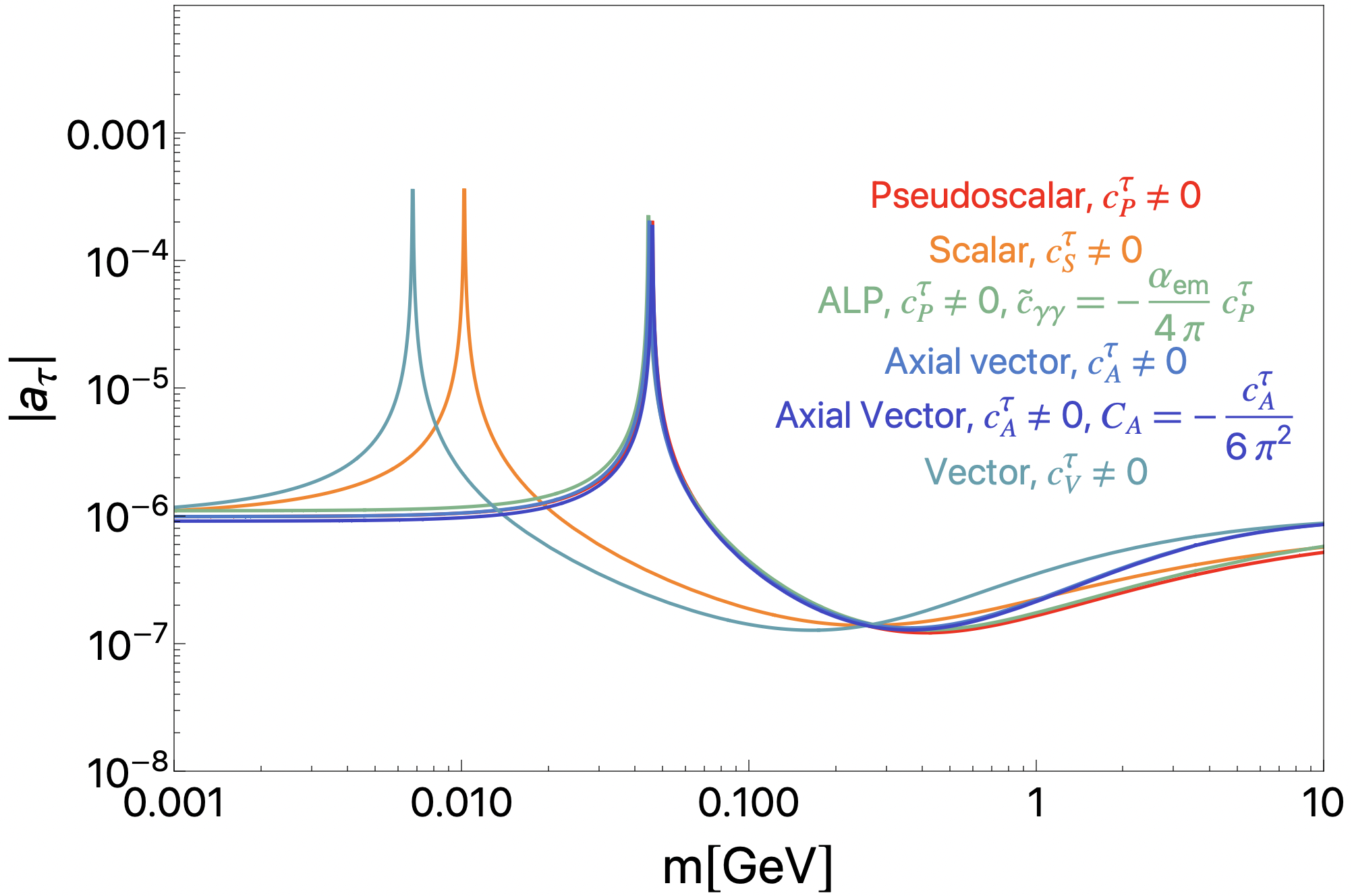}
    \includegraphics[width=0.49\linewidth]{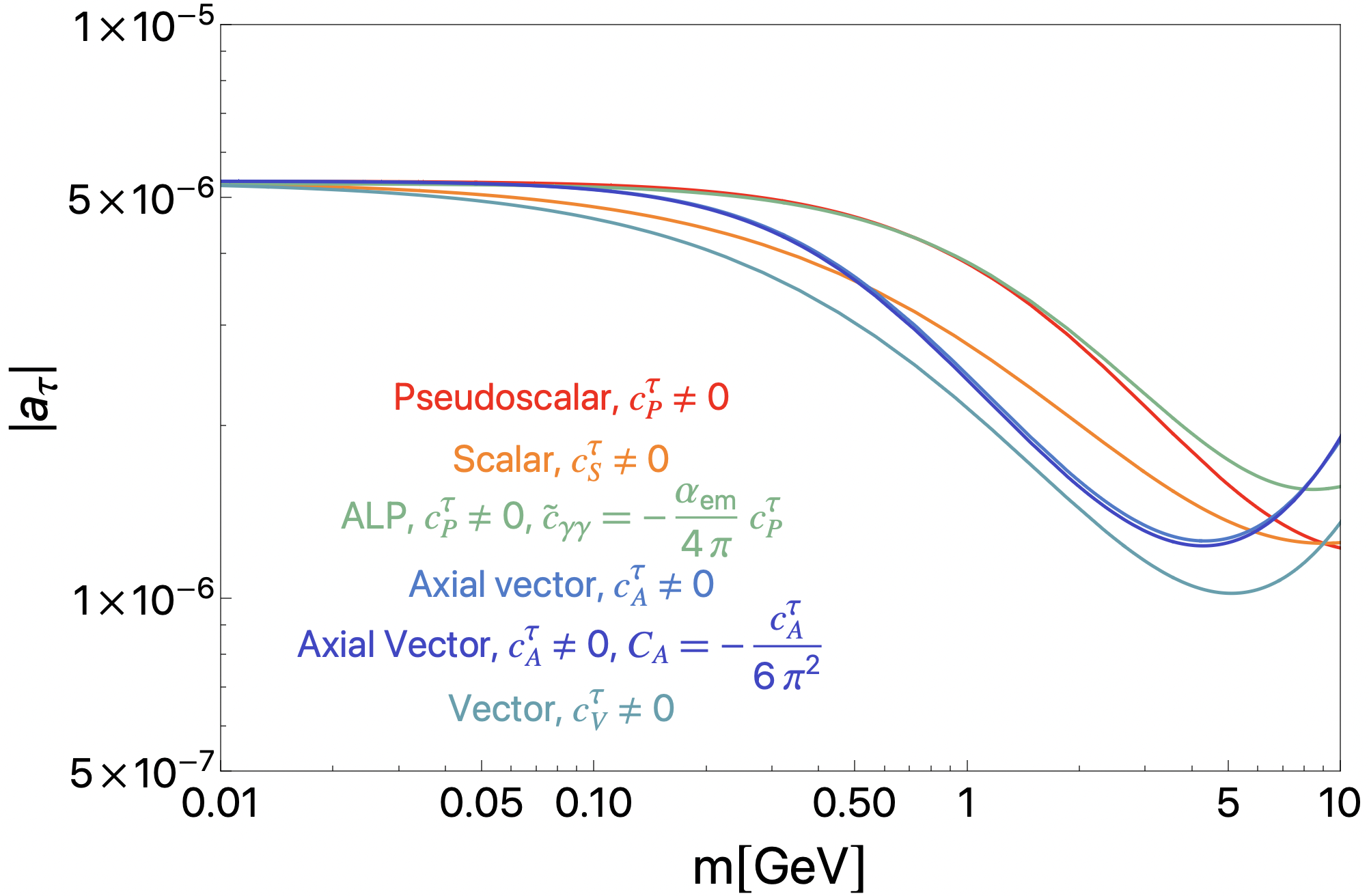}
    \caption{Comparison of the sensitivity of the Belle II experiment to light NP affecting $a_\tau$ just above the $\tau^+\tau^-$ threshold, $s_{\tau\tau} = 4(1.78\GeV)^2$ (same notation as in Fig.~\ref{fig:Money_Plots}). No loss in luminosity has been assumed for either of the two cases. Realistic projections should of course take it into account and rescale accordingly the results displayed here.}
    \label{fig:DiTau_threshold}
\end{figure*}

Assuming a given future sensitivity on $\Re F_{2,3}(s_B)$ and $\Im F_{2,3}(s_B)$ of $10^{-6}$, we can then estimate the sensitivity of Belle II to $a_\tau$ and $d_\tau$ at the CM energy $\sqrt{s_B}= 10.58\GeV$. Our key results are displayed in Fig.~\ref{fig:Money_Plots}. The first important observation is that due to $s_B>4m_\tau^2$, the form factors $F_{2,3}$ always develop an imaginary part. This is an important point that marks a fundamental difference between the light and the heavy NP case. Indeed, local, heavy NP (in the SMEFT sense) leaves a print exclusively on the real part of form factors, which can only be accessed provided that polarized electron beams are available and upon subtracting two asymmetries. On the other hand, if NP is light and generates nonlocal effects, it has an impact on the imaginary part of form factors. This class of effects is more easily accessible experimentally, as it does not require polarized beams and only relies on the measurement of the normal asymmetry.

Second, for large mediator masses the sensitivity to $a_\tau$ (and $d_\tau$) via $\Re F_{2,3}$ approaches a constant value, which coincides with the assumed sensitivity to $\Re F_{2,3}^\text{eff}$, in line with the EFT expectation. However, we observe that
scalar mediators display a slower decoupling as compared to vectors. This difference originates from a residual logarithmic dependence on the mass of the mediator, which is present in the scalar case but absent for vectors. Such a term is generated by the leftmost topology of diagrams in Fig.~\ref{fig:RepFeynmanDiagrams} and can again be understood from EFT arguments. Upon integrating out the heavy mediator, the diagram under consideration reduces to a diagram featuring a standard QED vertex and a four-fermion vertex. In the case of a vector such a four-fermion vertex is of the form $(\bar{L}L)(\bar{R}R)$, $(\bar{R}R)(\bar{R}R)$, or $(\bar{L}L)(\bar{L}L)$, while the scalar one gives rise to an $(\bar{L}R)(\bar{R}L)$ or an $(\bar{L}R)(\bar{L}R)$ structure. As a consequence of helicity selection rules~\cite{Cheung:2015aba}, only the latter can contribute to the renormalization of the effective dipole operator in LEFT~\cite{Alonso:2014rga, Jenkins:2017dyc}. The logarithmic term can then be interpreted as a renormalization effect, where the mass of the mediator signals the effective matching scale between the two regimes of the theory.

As a consequence of this decoupling behavior, spin-$1$ mediators fall back to the EFT limit much faster than their spin-$0$ counterparts. Moreover, including the dimension-5 nonrenormalizable interactions in Eq.~\eqref{eq:ALP_Lag} produces a case somewhere in between: in this case, the results
display a logarithmic sensitivity to the cutoff scale $\Lambda$ of the theory. As long as the mass of the mediator is symmetry-suppressed with respect to that scale, such a logarithmic dependence marks a behavior that differs from the $\log M_\phi^2$ dependence of the Yukawa-like interactions in Eq.~\eqref{eq:ALP_Lag}, but mimics a similar behavior once $M_\phi \to \Lambda$. In all cases we observe that the sensitivity to $a_\tau$ is lost whenever accidental cancellations occur, but elsewhere the typical sensitivity stays below $10^{-5}$ even for small mediator masses.

Finally, Fig.~\ref{fig:Money_Plots} also shows the sensitivity for $a_\tau$ that can be reached via $\Im F_2^\text{eff}$. That sensitivity disappears once the mediator mass is taken to infinity, as mandated by the EFT, but we observe the same decoupling behavior as before: the logarithmic terms for scalar mediators extend the sensitivity up to much higher mediator masses than for the vector case, and the nonrenormalizable example lies somewhere in between. The maximal sensitivity is reached around the CM energy $\sqrt{s_B}$, while for smaller mediator masses again a sensitivity below $10^{-5}$ remains.

In addition to our main results shown in Fig.~\ref{fig:Money_Plots}, obtained at a realistic Belle II CM energy, we also consider the case $s \simeq 4m_\tau^2$, which displays
 a near-threshold enhancement, see Fig.~\ref{fig:DiTau_threshold}. In principle,
 it is possible to take advantage of this behavior to further improve the experimental sensitivity on the quantities of interest up to at most one order of magnitude. Of course, such a  measurement would require a CM energy that differs significantly from the nominal Belle II operational one. Accessing such a regime would either require tuning the CM energy to threshold values, or making use of radiative return techniques. While the former is unlikely to happen, the latter could be applied at Belle II with the current experimental setup, of course with corresponding loss in statistics.

\emph{A specific example}---In order to further illustrate the points discussed above, we now consider a toy model featuring only Yukawa-like couplings to $\tau$ leptons, corresponding to the first line of Eq.~\eqref{eq:ALP_Lag}.\footnote{For an overview on $\tau$-philic scalars and axionlike particles, see Ref.~\cite{Alda:2024cxn}, while for $CP$-violating scalars and axionlike particles we refer to Ref.~\cite{DiLuzio:2023lmd}.}
In the limit $s \to 0$ one finds~\cite{Leveille:1977rc}
\begin{align}
\label{atau_Yukawa}
a_\tau &= \frac{m_\tau^2}{8\pi^2}\frac{m_\tau^2}{\Lambda^2} \int_0^1 dx \frac{(c_P^\tau)^2(2x^2-x^3)-x^3(c_S^\tau)^2}{m_\tau^2 x^2 + M_\phi^2(1-x)}\notag\\
&\simeq\frac{1}{16\pi^2}\frac{m_\tau^2}{M_\phi^2}\bigg[\left(\frac{m_\tau c_P^\tau}{\Lambda}\right)^2\left(\frac{11}{3}+ 4 \log\frac{m_\tau}{M_\phi}\right)\notag\\
& \qquad \qquad -\left(\frac{m_\tau c_S^\tau }{\Lambda}\right)^2\left(\frac{7}{3}+ 4 \log \frac{m_\tau}{M_\phi}\right)\bigg]\,,
\end{align}
where we kept the leading terms for large mediator mass. In the same limit, $M_\phi\to\infty$, we obtain
\begin{align}
\label{F2_Yukawa}
F_2(s) &\simeq \frac{1}{48\pi^2}\frac{m_\tau^2}{M_\phi^2} \bigg[\left(\frac{c_{P}^\tau m_\tau}{\Lambda}\right)^2\left(-1-6 \log\frac{M_\phi^2}{-s}\right) \notag \\
& \qquad \qquad +  \,\left(\frac{c_{S}^\tau m_\tau}{\Lambda}\right)^2\left(5+6 \log\frac{M_\phi^2}{-s}\right)\bigg] \,.
\end{align}
From these expressions one can read off key features of our numerical findings above.
 In the large-mediator-mass limit, the ratio $a_\tau/F_2$ differs from $1$ due to both finite terms and ratios of logarithms. The former can be understood as stemming from a different hierarchy in the limiting procedure, $M_\phi^2 \gg s> 4m_\tau^2$ for $F_2$ and $M_\phi^2 \gg m_\tau^2 \gg s$ in the case of $a_\tau$, while the appearance of the logarithms is a consequence of the EFT arguments given above. From the explicit expressions in Eqs.~\eqref{atau_Yukawa} and~\eqref{F2_Yukawa} one can check that indeed the coefficients of $\log M_\phi$ agree, thus ensuring the decoupling albeit only up to logarithmic corrections.

In fact, the exact same behavior, i.e., a leading-order logarithmic enhancement and corresponding decoupling pattern of a scalar mediator, already occurs in the SM.
The Higgs contribution to $a_\tau$ behaves as~\cite{Jackiw:1972jz,Bars:1972pe, Altarelli:1972nc,Bardeen:1972vi,Fujikawa:1972fe}
\begin{equation}
 a_\tau^h \propto \frac{m_\tau^2}{v^2} \frac{m_\tau^2}{M_h^2}\log \frac{M_h^2}{m_\tau^2}
 =\frac{m_\tau^4}{2\lambda v^4}\log \frac{2\lambda v^2}{m_\tau^2}\,,
\end{equation}
where $v$ is the electroweak vacuum expectation value and $\lambda$ is the Higgs quartic coupling.
In our scenario, the same result is found, provided that one replaces $v \to \Lambda$, $2 \lambda v = M_h \to M_\phi = 2\tilde{\lambda} \Lambda$. $\lambda$ and $\tilde{\lambda}$ are dimensionless constants that quantify the hierarchy between the symmetry breaking scale ($v$ for the electroweak symmetry and $\Lambda$ for NP) and the mass of the related bosonic mediator. If the mass of the latter is generated by a mechanism that respects the symmetry, such a constant is ${\mathcal O}(1)$ (as for the Higgs), while if the mass term explicitly breaks the symmetry associated to the cutoff scale, it is necessarily smaller (as, e.g., for a Goldstone boson). In case no large separation of scales is present within the NP sector, i.e., $\tilde{\lambda}\simeq 1$, or equivalently for large mediator masses, the SM behavior is reproduced, and a leading-order logarithmic behavior in $\Re F^\tau_2/a_\tau \simeq m_\tau^2/M_\phi^2\log [M_\phi^2/m_\tau^2]$ is induced.
As an aside, we also observe that the Higgs-mediated one-loop contributions to $a_\tau$ naturally develop an imaginary part. The magnitude of the latter can be estimated to be $\Im F_2^h \simeq  m_\tau^4/(8\pi M_h^2 v^2)\simeq 10^{-9}$. Any modification to our results due to such a term would fall by orders of magnitude beyond the projected sensitivity of $F_2 \simeq 10^{-6}$ and therefore it can be safely neglected.

\emph{Conclusions}---In this Letter we explored the possibility of probing the anomalous magnetic moment and the electric dipole moment of the $\tau$ lepton at Belle II via asymmetry measurements in $e^+e^-\to\tau^+\tau^-$ in the case of light NP. While for heavy mediators a general EFT argument ensures that measurements of the effective form factors $\Re F_{2,3}^\text{eff}$ can be interpreted directly as NP contributions $a_\tau^\text{NP}$ and $d_\tau^\text{NP}$, for light mediators the limits become model dependent. We performed the analysis for a representative set of spin-$0$ and spin-$1$ NP scenarios, finding a similar pattern in all cases, in that apart from accidental cancellations the NP sensitivity does not deteriorate by more than one order of magnitude compared to the EFT limit. In the case of spin-$0$ mediators we further observed an interesting decoupling pattern, in that the EFT limit is only approached logarithmically, mirroring a similar behavior of the Higgs contribution in the SM.

In addition to this sensitivity study, we also explored novel avenues for NP searches that become possible for light mediators. First, one can profit from threshold enhancement when tuning the CM energy close to $\sqrt{s}=2m_\tau$, which could potentially be accessed in radiative-return mode. More strikingly, the imaginary part developed above the $\tau^+\tau^-$ threshold provides another avenue for extracting limits on $a_\tau$, which is even possible in the current setting at Belle II without the need for a polarized electron beam.
While eventually the sensitivity disappears in the EFT limit, competitive limits can be obtained over a wide range of masses especially in the case of a spin-$0$ mediator. A measurement of the normal asymmetry
would thus not only demonstrate the method, but at the same time already provide interesting physics insights in its own right,
strongly motivating such a measurement at Belle II. With sufficient statistics, a similar program could be imagined at other $e^+e^-$ machines, such as BESIII~\cite{BESIII:2020nme} or the proposed Super Tau-Charm Factory~\cite{Achasov:2023gey}.

 \section*{Acknowledgments}
Financial support by the SNSF (Project No.\ TMCG-2\_213690) is gratefully acknowledged.

\bibliography{tau}

\end{document}